\documentclass{SciPost}

\hypersetup{
    colorlinks,
    linkcolor={red!50!black},
    citecolor={blue!50!black},
    urlcolor={blue!80!black}
}

\usepackage[bitstream-charter]{mathdesign}
\usepackage{xspace}
\DeclareSymbolFont{usualmathcal}{OMS}{cmsy}{m}{n}
\DeclareSymbolFontAlphabet{\mathcal}{usualmathcal}

\fancypagestyle{SPstyle}{
\fancyhf{}
\lhead{\colorbox{scipostdeepblue}{\bf \color{white} ~SciPost Physics Proceedings }}
\rhead{{\bf \color{scipostdeepblue} ~Submission }}

\fancyfoot[C]{\textbf{\thepage}}
}

\newcommand{\aiinfn}{\ensuremath{\mathrm{AI\_INFN}}\xspace}

\newcommand\jhub{\ensuremath{\mathrm{JupyterHub}}\xspace}
\newcommand\jlab{\ensuremath{\mathrm{JupyterLab}}\xspace}
\newcommand\infncloud{\ensuremath{\mathrm{INFN\ Cloud}}\xspace}
\newcommand\datacloud{\ensuremath{\mathit{DataCloud}}\xspace}

\newcommand\ks{\ensuremath{\mathrm{Kubernetes}}\xspace}

\begin{document}

\pagestyle{SPstyle}

\begin{center}{\Large \textbf{\color{scipostdeepblue}{
The AI\_INFN Platform:\\Artificial Intelligence Development in the Cloud\\
}}}\end{center}

\begin{center}\textbf{
Lucio Anderlini\textsuperscript{1},
Giulio Bianchini\textsuperscript{2},
Diego Ciangottini\textsuperscript{2},
Stefano Dal Pra\textsuperscript{3},
Diego Michelotto\textsuperscript{3},
\textbf{Rosa Petrini}\textsuperscript{1$\star$} and
Daniele Spiga\textsuperscript{2}
}\end{center}

\begin{center}
{\bf 1} Istituto Nazionale di Fisica Nucleare (INFN), Sezione di Firenze, Italy
\\
{\bf 2} Istituto Nazionale di Fisica Nucleare (INFN), Sezione di Perugia, Italy
\\
{\bf 3} Istituto Nazionale di Fisica Nucleare (INFN), CNAF, Italy
\\[\baselineskip]
$\star$ \href{mailto:rosa.petrini@fi.infn.it}{\small rosa.petrini@fi.infn.it}
\end{center}

\definecolor{palegray}{gray}{0.95}
\begin{center}
\colorbox{palegray}{
  \begin{tabular}{rr}
  \begin{minipage}{0.37\textwidth}
    \includegraphics[width=60mm]{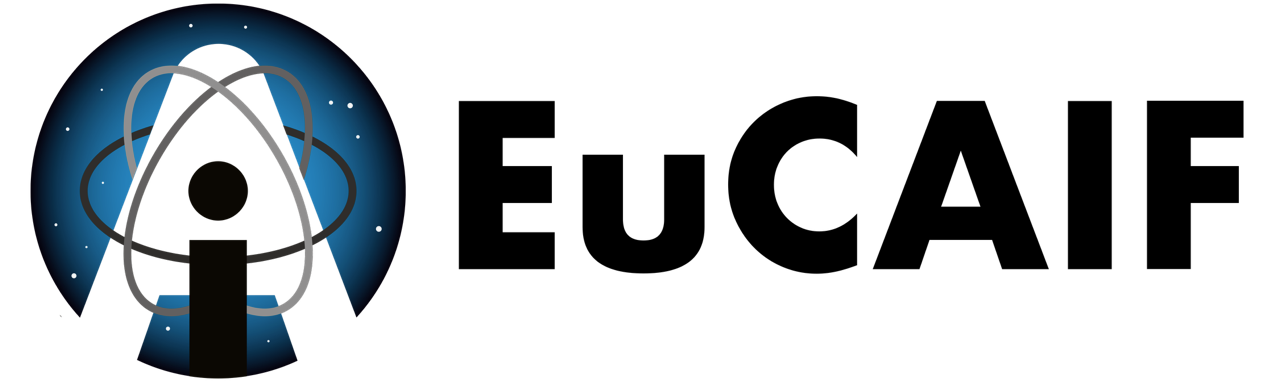}
  \end{minipage}
  &
  \begin{minipage}{0.5\textwidth}
    \vspace{5pt}
    \vspace{0.5\baselineskip} 
    \begin{center} \hspace{5pt}
    {\it The 2nd European AI for Fundamental \\Physics Conference (EuCAIFCon2025)} \\
    {\it Cagliari, Sardinia, 16-20 June 2025
    }
    \vspace{0.5\baselineskip} 
    \vspace{5pt}
    \end{center}
    
  \end{minipage}
\end{tabular}
}
\end{center}

\section*{\color{scipostdeepblue}{Abstract}}
\textbf{\boldmath{%
Machine Learning (ML) is profoundly reshaping the way researchers create, implement, and operate data-intensive software. Its adoption, however, introduces notable challenges for computing infrastructures, particularly when it comes to coordinating access to hardware accelerators across development, testing, and production environments.
The INFN initiative \aiinfn (Artificial Intelligence at INFN) seeks to promote the use of ML methods across various INFN research scenarios by offering comprehensive technical support, including access to AI-focused computational resources. Leveraging the INFN Cloud ecosystem and cloud-native technologies, the project emphasizes efficient sharing of accelerator hardware while maintaining the breadth of the Institute’s research activities.
This contribution describes the deployment and commissioning of a Kubernetes-based platform designed to simplify GPU-powered data analysis workflows and enable their scalable execution on heterogeneous distributed resources. By integrating offloading mechanisms through Virtual Kubelet and the InterLink API, the platform allows workflows to span multiple resource providers, from Worldwide LHC Computing Grid sites to high-performance computing centers like CINECA Leonardo. We will present preliminary benchmarks, functional tests, and case studies, demonstrating both performance and integration outcomes.
}}

\vspace{\baselineskip}

\noindent\textcolor{white!90!black}{%
\fbox{\parbox{0.975\linewidth}{%
\textcolor{white!40!black}{\begin{tabular}{lr}%
  \begin{minipage}{0.6\textwidth}%
    {\small Copyright attribution to authors. \newline
    This work is a submission to SciPost Phys. Proc. \newline
    License information to appear upon publication. \newline
    Publication information to appear upon publication.}
  \end{minipage} & \begin{minipage}{0.4\textwidth}
    {\small Received Date \newline Accepted Date \newline Published Date}%
  \end{minipage}
\end{tabular}}
}}
}




\section{Introduction}
The rapid expansion of Machine Learning (ML) and Artificial Intelligence (AI) is fundamentally transforming data-intensive scientific research. In domains like High Energy Physics (HEP), where experiments produce massive and intricate datasets, these advanced computational methods have become essential for tasks ranging from data collection and simulation to final physics analyses~\cite{Albertsson:2018maf}.
The Italian National Institute for Nuclear Physics (INFN), a leading institution in subnuclear, nuclear, and astroparticle physics, acknowledges this shift and is adapting its computing infrastructure to meet the growing demands of AI and cloud technologies. Through INFN Cloud~\cite{Salomoni:2024dft,infn_cloud}, resources~\cite{Grandi:2024fqt} are being upgraded and services broadened to support this transition.
Within this framework, the \aiinfn initiative has been launched to integrate AI-driven technologies across INFN’s research areas. Its primary goal is to build a unified ecosystem connecting infrastructure, algorithms, and applications, fostering collaboration, and accelerating scientific discoveries. The initiative is structured around four main work packages:
\begin{enumerate}
    \item Infrastructure Provisioning: Establishing and operating advanced computing infrastructure equipped with GPUs and other hardware accelerators.
    \item Training and Education: Hosting workshops, hackathons, and training sessions to promote proficiency in modern ML techniques.
    \item Community Building: Developing a network of ML practitioners and developers across INFN units to facilitate knowledge sharing and best practices.
    \item Future Technologies: Exploring next-generation hardware for AI applications, including FPGAs and quantum processors, to enhance model training and inference.
\end{enumerate}
The \aiinfn platform~\cite{Anderlini:chep2025,WMLQ2024}, developed in partnership with the \datacloud initiative, operating \infncloud, provides a cloud-native, scalable, and user-focused environment. It empowers INFN researchers with the necessary tools to advance AI-powered scientific exploration and support the development of innovative workflows and analyses.

\section{The \aiinfn Platform Architecture}
The transition to the \aiinfn platform represents a strategic shift from a rigid, resource-centric infrastructure to a flexible, cloud-native, service-oriented architecture. This change addresses key limitations of the previous VM-based model~\cite{Anderlini:mlinfn2024}, such as inefficient use of accelerators, risks of data loss, and unsustainable administrative and security demands. To overcome these obstacles, the \aiinfn platform was redesigned as a \textit{Software As A Service} (SaaS) application built on \ks~\cite{K8s}. This modern, cloud-native approach decouples computing resources from user data, enabling dynamic hardware allocation without compromising data integrity. A typical example 
is \jhub~\cite{jupyter}, which provides multiple users with access to notebooks for data visualization and interactive computation using resources provisioned by the Cloud. This strategic pivot to a \ks-based SaaS model resolves the issues of static allocation and data fragility. The following section details the components of the new architecture, from the hardware layer to user-facing services, that enable flexibility and scalability.

At its core, the \aiinfn platform is a managed \ks cluster that abstracts the complexity of its underlying high-performance hardware. This allows researchers to focus on their research rather than on infrastructure. The platform's architecture integrates high-performance hardware, a flexible software stack, and essential services that empower researchers to develop, test, and scale their AI-driven workflows.

The hardware is hosted at the INFN CNAF data center in Bologna and consists of four high-performance servers, clustered in an OpenStack \cite{OpenStack} tenancy, acquired between 2020 and 2024 to meet increasing computational demands.

\begin{itemize}
    \item Server 1 (2020): 64 CPU cores, 750 GB of memory, 12 TB of NVMe disk, 8 NVIDIA Tesla T4 GPUs and 5 NVIDIA RTX 5000 GPUs;
    \item Server 2 (2021): 128 CPU cores, 1024 GB of memory, 12 TB of NVMe disk, 2 NVIDIA Ampere A100 GPU, 1 NVIDIA Ampere A30 GPU, 2 AMD-Xilinx U50 boards and 1 AMD-Xilinx U250 board;
    \item  Server 3 (2023): 128 CPU cores, 1024 GB of memory, 24 TB of NVMe disk, 3 NVIDIA Ampere A100 GPUs and 5 AMD-Xilinx U250 boards;
    \item Server 4 (2024): 128 CPU cores, 1024 GB of memory, 12 TB of NVMe disk, 1 NVIDIA RTX 5000 GPUs and 2 AMD-Xilinx Alveo U55c.
\end{itemize}

The main platform file system is distributed through the containers via NFS. One of the platform nodes runs an NFS server in a \ks pod and exports data to the containers spawned by \jhub.
At spawn time, \jhub is configured to create the user's home directories and project-dedicated shared volumes. 
A special directory of the platform file system, that users can use directly or clone and extend in their directories, is reserved for distributing managed software environments, configure using virtual environments.
The platform file system is subject to regular encrypted backup. 
Backup data is stored in a remote Ceph volume~\cite{ceph} provisioned  by \infncloud using the \emph{BorgBackup}~\cite{borg} package to ensure data deduplication.

Large datasets must be stored in a centralized object storage service based on Rados Gateway~\cite{radosgw} and centrally managed by \datacloud.
To ease accessing the datasets with the Python frameworks commonly adopted in Machine Learning projects, a patched version of \texttt{rclone}~\cite{rclone} was developed to enable mounting the user's bucket in the \jlab instance using the same authentication token used to access \jhub. The mount operation is automated at spawn time.

Efficient GPU management is achieved through the adoption of the NVIDIA GPU Operator~\cite{NVIDIA-GPU-Op} for \ks. This tool automates the management of GPU driver lifecycles, container runtimes, and monitoring components. It simplifies administration and ensures consistent configurations across the cluster. One notable benefit is the ability to partition NVIDIA A100 GPUs into multiple, independent Multi-Instance GPUs~\cite{MIG} (MIGs). This feature enables a single physical GPU to serve up to seven users simultaneously, significantly increasing access to high-demand accelerator resources. The operator-based approach centralizes and standardizes GPU lifecycle management, reducing administrative overhead significantly compared to maintaining custom configurations on individual VMs.

The platform supports multiple mechanisms to provide customizable software environments. Users can start with preconfigured, templated Conda~\cite{conda} environments or Apptainer~\cite{Apptainer} images for common frameworks, such as TensorFlow~\cite{tensorflow}, Torch~\cite{torch}, and Keras~\cite{keras}. Specialized environments for emerging fields such as Quantum Machine Learning (QML) are also available. For maximum flexibility, users can deploy custom OCI (Open Container Initiative) images, allowing them to freely define the system libraries and software packages required for their research.

A dedicated monitoring and accounting system provides crucial visibility into resource utilization. The system uses Prometheus~\cite{prometheus} to collect metrics from various exporters, including Kube Eagle~\cite{kubeeagle} for CPU/memory usage and the NVIDIA DCGM Exporter~\cite{dcgmexporter} for GPU telemetry. In addition to off-the-shelf exporters, custom exporters were developed to monitor specific resources, such as storage utilization. These metrics are visualized through Grafana~\cite{grafana} dashboards, enabling administrators to effectively control resource allocation and plan for future capacity needs. This robust architecture provides a comprehensive solution for interactive development and serves as the foundation for the platform's advanced scaling capabilities. At the same time, a feasibility study is ongoing to develop personalized user dashboards,
so that each researcher can monitor and manage their own resource consumption. This feature is expected to significantly enhance transparency and foster responsible usage of the shared infrastructure.

At the time of writing, 78 \infncloud users registered to the \aiinfn platform and 20 multi-user research projects were allocated.

\section{Batch Processing and Workload Offloading}

Seamlessly scaling workloads from interactive development to large-scale batch processing is a critical requirement for modern scientific computing. The \aiinfn platform addresses this with a two-pronged strategy: an opportunistic local batch system and a sophisticated offloading architecture that connects the platform with national-level HPC assets like the CINECA Leonardo supercomputer~\cite{Leonardo}.

The local batch system is managed by Kueue~\cite{Kueue}, a \ks-native job queue controller. It is designed to opportunistically run non-interactive workloads, making effective use of the cluster’s resources during off-peak hours, such as nights and weekends. To ensure a responsive experience for interactive users, Kueue is configured to prioritize \jlab sessions. If resource contention occurs, running batch jobs are automatically evicted to free up hardware for interactive development.

In the field of workflow definition, Snakemake has emerged as a promising infrastructural component. Providing an alternative to traditional Job Description Languages, it offers explicit handling of job dependencies and reproducible workflows. Snakemake workflows can be entirely submitted to the platform, where job dependencies are managed by a dedicated controller. For workloads that exceed the local cluster's capacity, the platform features an offloading architecture that transparently executes jobs on external computing resources. Virtual Kubelet~\cite{virtualkubelet} enables this by allowing a Kubernetes cluster to treat a remote resource provider as if it were a local node. The \aiinfn platform relies on the InterLink~\cite{Ciangottini:chep2024} provider. 
Successful scalability tests have validated this architecture by orchestrating workloads across four different sites using heterogeneous schedulers (HTCondor~\cite{htcondor} and SLURM~\cite{slurm}) and backends (Podman~\cite{podman}). These tests integrated resources from the INFN-Tier1 at CNAF, ReCaS Bari and the CINECA Leonardo supercomputer, providing powerful proof of the platform’s ability to manage complex, federated workflows and prove its readiness to tackle demanding computational challenges in the highly heterogeneous landscape of the Italian computing ecosystem.

\section{Conclusion}

The \aiinfn platform represents a significant advancement in the computing infrastructure available to the Italian scientific community. By evolving from a static, VM-based model to a flexible, cloud-native \textit{Software as a Service} (SaaS) environment, the platform successfully addresses the core challenges of modern scientific computing. It provides researchers with a highly customizable, scalable, collaborative platform that simplifies access to high-performance hardware accelerators and modern software tools. This environment fosters the adoption of cutting-edge machine learning techniques across a wide range of INFN research domains, including high-energy physics, solid-state detector modeling, theoretical physics, and nuclear medicine.
Although initially conceived as a research initiative, the \aiinfn platform is quickly emerging as a key provider of GPU-accelerated computing resources. By empowering scientists to extract insights from complex data, it is positioned to play a pioneering role in shaping the future of INFN’s computing infrastructure.

\section*{Acknowledgements}
 
\textbf{Funding information} The work presented in this paper, and in particular the work of G.B., D.C. and R.P., has been funded in the framework of Spoke~0 and Spoke~2 of the ICSC project -- \emph{Centro Nazionale 
di Ricerca in High Performance Computing, Big Data and Quantum 
Computing}, funded by the NextGenerationEU European initiative through 
the Italian Ministry of University and Research, PNRR Mission~4, 
Component~2: Investment~1.4, Project code CN00000013 - CUP 
I53C21000340006.



\bibliography{bibliography.bib}


\end{document}